\documentclass[10pt,conference]{IEEEtran}
\usepackage{graphicx}
\usepackage{epsfig}
\usepackage{amsmath,amssymb,amsfonts,cite}

\def\Pr{\mathrm{Pr}}

\def\g{{\mathrm g}}

\def\Rc{\mathcal R}
\def\Xc{\mathcal X}
\def\Yc{\mathcal Y}
\def\Wc{\mathcal W}
\def\Cc{\mathcal C}

\newtheorem{theorem}{Theorem}

\newtheorem{remark}{Remark}

\IEEEoverridecommandlockouts

\begin{document}

\title{Interference-Assisted Secret Communication}

%
\author{\authorblockN{Xiaojun Tang\authorrefmark{1},
Ruoheng Liu\authorrefmark{2}, Predrag
Spasojevi\'{c}\authorrefmark{1}, and H. Vincent
Poor\authorrefmark{2}}
\thanks{This research was supported by the National Science Foundation under Grants ANI-03-38807, CNS-06-25637 and CCF-07-28208.}
\thanks{${\ast}$ X. Tang and P. Spasojevi\'{c} are with Wireless Information Network Laboratory (WINLAB), Department of Electrical and Computer Engineering, Rutgers University,
North Brunswick, NJ 08902, USA (e-mail:
\{xtang,spasojev\}@winlab.rutgers.edu).}%
\thanks{${\dag}$ R. Liu and H. V. Poor are with Department of Electrical Engineering, Princeton
University, Princeton, NJ 08544, USA (email:
\{rliu,poor\}@princeton.edu).} }


\maketitle

\begin{abstract}
Wireless communication is susceptible to adversarial
eavesdropping due to the broadcast nature of the wireless
medium. In this paper it is shown how eavesdropping can be
alleviated by exploiting the superposition property of the wireless
medium. A wiretap channel with a helping interferer (WT-HI), in
which a transmitter sends a confidential message to its intended
receiver in the presence of a passive eavesdropper, and with the
help of an independent interferer, is considered. The interferer,
which does not know the confidential message, helps in ensuring
the secrecy of the message by sending independent signals. An
achievable secrecy rate for the WT-HI is given. The results show
that interference can be exploited to assist secrecy in wireless
communications. An important example of the Gaussian case,
in which the interferer has a better channel to the intended
receiver than to the eavesdropper, is considered. In this
situation, the interferer can send a (random) codeword at a rate
that ensures that it can be decoded and subtracted from the
received signal by the intended receiver but cannot be decoded
by the eavesdropper. Hence, only the eavesdropper is interfered
with and the secrecy level of the confidential message is increased.
\end{abstract}

\section{Introduction}\label{sec:intro}

Broadcast and superposition are two fundamental properties of the
wireless medium. Due to the broadcast nature, wireless transmission
can be received by multiple receivers with possibly different signal
strengths. Due to the superposition property, a receiver observes a
signal that is a superposition of multiple simultaneous
transmissions. From the \textit{secure communication} point of view,
both features pose a number of security issues. In particular, the
broadcast nature makes wireless transmission susceptible
to \textit{eavesdropping}, because anyone (including adversarial
users) within the communication range can listen and possibly
extract the confidential information. The superposition property
makes wireless communication susceptible to \textit{jamming}
attacks, where adversarial users can superpose destructive signals
(interference) onto useful signals to block the intended
transmission.

A helper can pit one property of the wireless medium against the
security issues caused by the other. An example in which broadcast is
employed to counteract the effects of superposition is the case of a
helper that functions as a relay to facilitate the transmission
from a source terminal to a severely jammed destination terminal. In
this paper, we consider the case in which a helper functions as an
\textit{interferer} to improve the secrecy level of a communication
session which is compromised by a passive eavesdropper. This is an
example where superposition is employed to counteract the security
threat due to the broadcast nature of the wireless medium.

We study the problem in which a transmitter sends confidential
messages to an intended receiver with the help of an interferer, in
the presence of a passive eavesdropper. We call this model the
\textit{wiretap channel with a helping interferer} (WT-HI for
brevity). In this system, it is desirable to minimize the leakage of
information to the eavesdropper. The interferer tries to help by
transmitting a signal without knowledge of the actual
confidential message. The level of ignorance of the eavesdropper
with respect to the confidential messages is measured by the
equivocation rate. This information-theoretic approach was
introduced by Wyner for the \textit{wiretap channel}
\cite{Wyner:BSTJ:75}, in which a single source-destination
communication is eavesdropped upon via a degraded channel. Wyner's
formulation was generalized by Csisz{\'{a}}r and K{\"{o}}rner who
determined the capacity region of the broadcast channel with
confidential messages \cite{Csiszar:IT:78}. The Gaussian wiretap
channel was considered in \cite{Leung-Yan-Cheong:IT:78}. More
recently, there has been a resurgence of interest in
\textit{information-theoretic security} for multi-user channel
models. Related prior work includes the multiple access channel (MAC) with
confidential
messages\cite{Liang:IT:06,Liu:ISIT:06,Tekin:IT:06,Tang:ITW:07,Tekin:IT:07},
the interference channel with confidential messages \cite{Liu:IT:07,
Liang:Allerton:07}, and the relay-eavesdropper channel
\cite{Lai:IT:06,Yusel:CISS:07}.

In this paper, an achievable secrecy rate for the WT-HI under the
requirement of \textit{perfect secrecy} is given. That is, the
eavesdropper is kept in total ignorance with respect to the message
for the intended receiver. A geometrical interpretation of the achievable
secrecy rate is given based on the MAC achievable rate regions from the
transmitter and the interferer to the intended receiver and to the
eavesdropper, respectively. For a symmetric Gaussian WT-HI, both the
achievable secrecy rate and a power control scheme are given. The
results show that the interferer can increase the secrecy level,
and that a positive secrecy rate can be achieved even when the
source-destination channel is worse than the source-eavesdropper
channel. An important example of the Gaussian case is that in which
the interferer has a better channel to the intended receiver than to
the eavesdropper. Here, the interferer can send a (random) codeword
at a rate that ensures that it can be decoded and subtracted from
the received signal by the intended receiver, but cannot be decoded by the
eavesdropper. Hence, only the eavesdropper is interfered with and
the secrecy level of the confidential message can be increased. Our
scheme can be considered to be a generalization of the two schemes
in [8], [9], and [11]. In the cooperative jamming [8] (artificial noise [9]) scheme, the helper generates an independent Gaussian noise. This
scheme does not employ any structure in the transmitted signal. The
noise forwarding scheme in [11] requires that the interferer's
codewords can always be decoded by the intended receiver, which is
not necessary in our scheme.

The remainder of the paper is organized as follows.
Section~\ref{sec:model} describes the system model for the WT-HI.
Section \ref{sec:result} states an achievable secrecy rate followed
by its geometrical interpretations in Section \ref{sec:GI}. Section
\ref{sec:Gaussian} gives the achievable secrecy rate and a power
control scheme for a symmetric Gaussian WT-HI. Section
\ref{sec:numerical} illustrates the results through some numerical
examples. Conclusions are given in Section~\ref{sec:conclusions}.

\section{System Model}\label{sec:model}

We consider a communication system including a transmitter ($X_1$),
an intended receiver ($Y_1$), a helping interferer ($X_2$), and a
passive eavesdropper ($Y_2$). The transmitter sends a confidential
message $W$ to the intended receiver with the help from an
\textit{independent} interferer, in the presence of a passive but
\textit{intelligent} eavesdropper. We assume that the helper does
not know the confidential message $W$ and the eavesdropper knows
codebooks of the transmitter and helper. As noted above, we refer to
this channel as the wiretap channel with a helping-interferer (WT-HI).
The channel can be defined by the alphabets $\Xc_1$, $\Xc_2$,
$\Yc_1$, $\Yc_2$, and channel transition probability
$p(y_1,y_2|x_1,x_2)$ where $x_t\in\Xc_t$ and $y_t\in\Yc_t$, $t=1,2$.

The transmitter uses encoder 1 to encode a confidential message $w
\in \Wc = \{1,\dots, M\}$ into $x_1^n$ and sends it to the intended
receiver in $n$ channel uses. A stochastic encoder
\cite{Csiszar:IT:78} $f$ is specified by a matrix of conditional
probabilities $f(x_{1,k}|w)$, where $x_{1,k} \in \Xc_1$, $w \in
\Wc$, $\sum_{x_{1,k}}f_1(x_{1,k}|w)=1$ for all $k=1,\dots, n$, and
$f(x_{1,k}|w)$ is the probability that encoder 1 outputs $x_{1,k}$
when message $w$ is being sent. The helper generates its output
$x_{2,k}$ randomly and can be considered as using another stochastic
encoder $f_2$, which is specified by a matrix of probabilities
$f_{2}(x_{2,k})$ with $x_{2,k} \in \Xc_{2}$ and
$\sum_{x_{2,k}}f_{2}(x_{2,k})=1.$  Since randomization can increase
secrecy, encoder 1 uses stochastic encoding to introduce
\textit{randomness}. Additional randomization is provided by the
helper and the secrecy can be increased further. 

The decoder uses the output sequence $y_1^n$ to compute its estimate
$\hat{w}$ of $w$. The decoding function is specified by a
(deterministic) mapping $g: \Yc_1^n \rightarrow \Wc$.

The average probability of error is
\begin{equation}\label{pe}
    P_e=\frac{1}{M}\sum_{w}\Pr\left\{g(Y_1^n) \neq w | w
    ~\mbox{sent}\right\}.
\end{equation}
The secrecy level (level of ignorance of the eavesdropper with
respect to the confidential message $w$) is measured by the
equivocation rate $(1/n)H(W|Y_2^n)$.

A secrecy rate $R_s$ is achievable for the WT-HI if, for any
$\epsilon>0$, there exists an ($M,n,P_e$) code so that

\begin{equation}\label{ach_def1}
    M \geq 2^{nR_s}, ~ P_e \leq \epsilon
\end{equation}
\begin{equation}\label{ach_def2}
\text{and} \qquad  R_s - \frac{1}{n}H(W|Z^n) \leq \epsilon \quad
\qquad ~
\end{equation}
for all sufficiently large $n$. The secrecy capacity is the maximal
achievable secrecy rate.

\section{Achievable Secrecy Rate}\label{sec:result}

\begin{theorem} \label{thm:WT-HI}
Let $\Rc_1$ denote the achievable rate region of the MAC $(\Xc_1,\Xc_2) \rightarrow \Yc_1$:
\begin{align}
\Rc_1^{[\rm MAC]}=\left\{(R_1,R_2)\left|
        \begin{array}{l}
          R_1\ge 0,~R_2\ge 0,\\
          R_1\le I(X_1;Y_1|X_2), \\
          R_2\le I(X_2;Y_1|X_1),\\
          R_1+R_2 \le I(X_1,X_2; Y_1)
        \end{array}
      \right.\right\}
\end{align}
and $\Rc_2$ denote the region of the MAC $(\Xc_1,\Xc_2) \rightarrow
\Yc_2$:
\begin{align}
\Rc_2^{[\rm MAC]}=\left\{(R_1,R_2)\left|
        \begin{array}{l}
          R_1\ge 0,~R_2\ge 0,\\
          R_1< I(X_1;Y_2|X_2), \\
          R_2< I(X_2;Y_2|X_1),\\
          R_1+R_2 < I(X_1,X_2; Y_2)
        \end{array}
      \right.\right\}.
\end{align}
We also define
\begin{align}
&   & \Rc_1^{[\rm S]}&=\left\{(R_1,R_2)\left|~
        \begin{array}{l}
          R_1\ge 0,~R_2\ge 0,\\
          R_1\le I(X_1;Y_1),\\
          R_2 >I(X_2;Y_1|X_1)
        \end{array}
      \right.\right\}\\
&\text{and}& \Rc_2^{[\rm S]}&=\left\{(R_1,R_2)\left|~
        \begin{array}{l}
          R_1\ge 0,~R_2\ge 0,\\
          R_1< I(X_1;Y_2),\\
          R_2 >I(X_2;Y_2|X_1)
        \end{array}
      \right.\right\}.
\end{align}
The following secrecy rate is achievable for the WT-HI:
\begin{align}
R_s=\max_{\pi, R_1,R_2,R_{1,d}}\left\{R_{1,s}\left|
        \begin{array}{l}
          R_{1,s}+R_{1,d}=R_1,\\
          (R_1,R_2)\in \left\{\Rc_1^{[\rm MAC]} \cup \Rc_1^{[\rm S]}\right\}, \\
          (R_{1,d},R_2) \notin \left\{\Rc_2^{[\rm MAC]} \cup \Rc_2^{[\rm S]}\right\}
        \end{array}
      \right.\right\},
\end{align}
where $\pi$ is the class of distributions that factor as
\begin{equation} \label{eq:dis-IC}
p(x_1)p(x_2)p(y_1,y_2|x_1,x_2).
\end{equation}
\end{theorem}
\begin{proof}
We briefly outline the achievable coding scheme here and omit the
details of the proof, which can be found in \cite{Tang:Preprint:07}.
We consider two independent stochastic codebooks. Encoder 1 uses
codebook $\Cc_1(2^{nR_1},2^{nR_{1,s}}, n)$, where $n$ is the
codeword length, $2^{nR_1}$ is the size of the codebook, and
$2^{nR_{1,s}}$ is the number of confidential messages that $\Cc_1$
can convey ($R_{1,s}\leq R$). In addition, encoder 2 uses codebook
$\Cc_2(2^{nR_2},n)$, where $2^{nR_2}$ is the codebook size. The
$2^{nR_1}$ codewords in codebook $\Cc_1$ are randomly grouped into
$2^{nR_{1,s}}$ bins each with $M=2^{n(R_1-R_{1,s})}$ codewords.
During the encoding, to send message $w \in [1,\dots,2^{nR_{1,s}}]$,
encoder 1 randomly selects a codeword from bin $w$ and sends to
channel, while encoder 2 randomly selects a codeword from codebook
$\Cc_2$ to transmit.
\end{proof}

\begin{remark}
The rate $R_1$ is split as $R_1 = R_{1,s} + R_{1,d}$, where
$R_{1,s}$ denotes a secrecy information rate intended by receiver 1
and $R_{1,d}$ represent a redundancy rate sacrificed in order to
confuse the eavesdropper. The interferer helps the receiver~1
confuse the eavesdropper by transmitting dummy information with rate
$R_2$.
\end{remark}

\section{Geometric Interpretations}\label{sec:GI}

When the intended receiver needs to decode both codewords from
$\Cc_1$ and $\Cc_2$, we essentially have a compound MAC.
\begin{figure}[htbp]
  \centerline{\hbox{ \hspace{0.01in}
    \epsfig{file=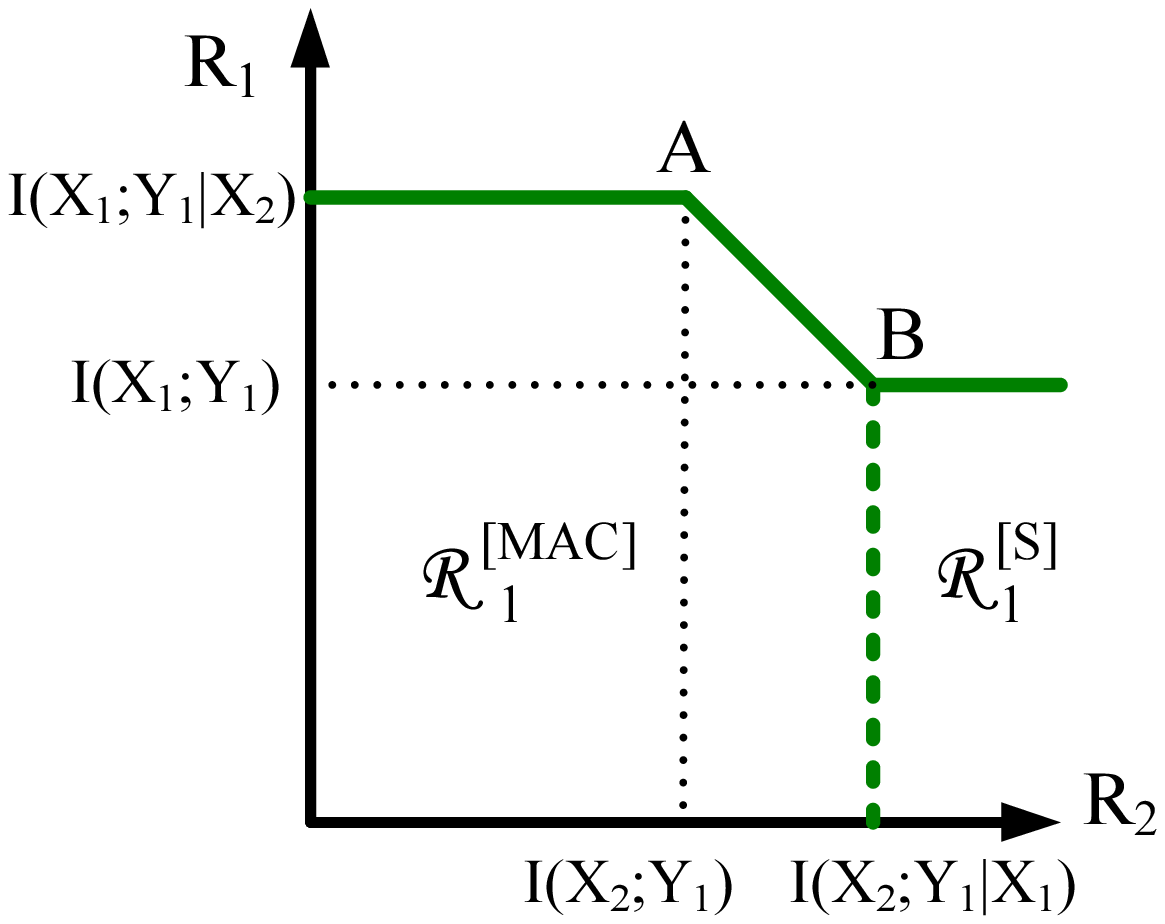, angle=0, width=0.2\textwidth}
    \hspace{0.01in}
    \epsfig{file=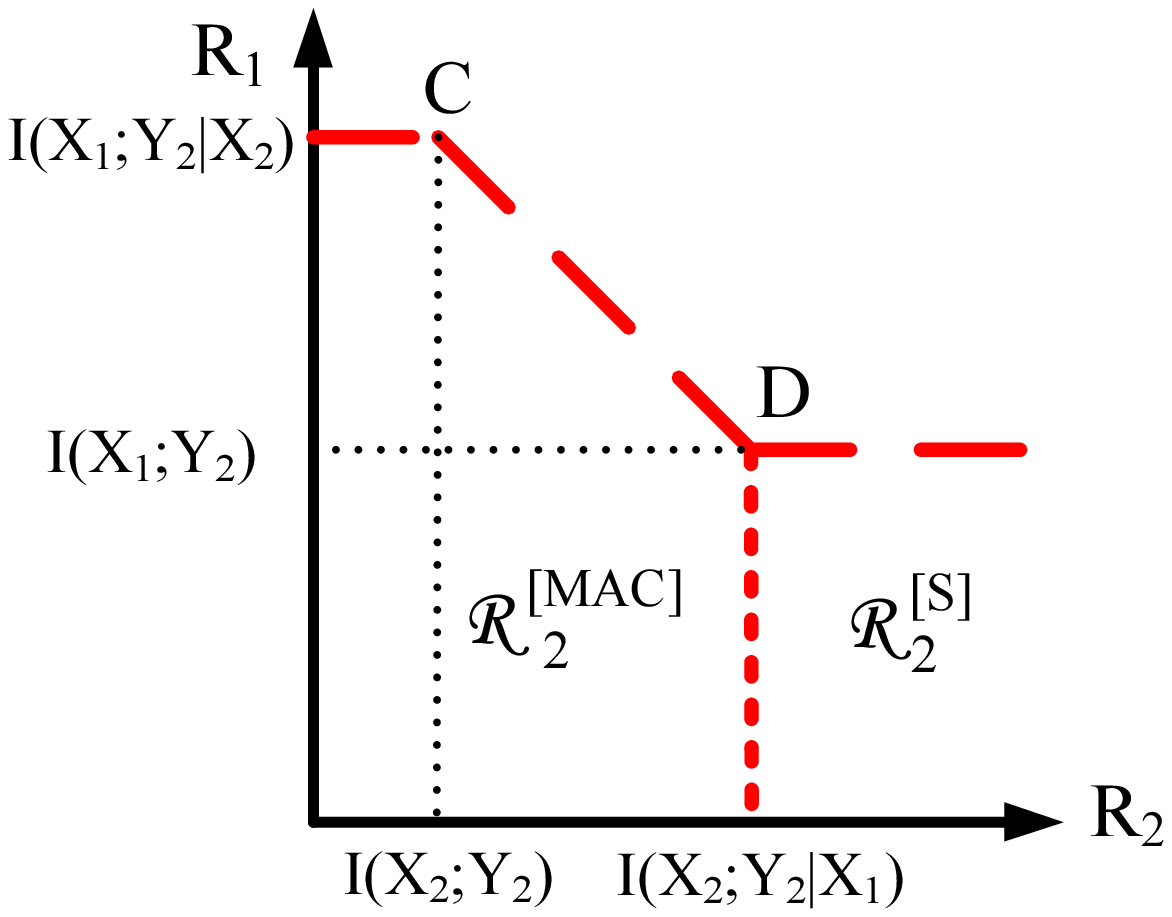, angle=0, width=0.2\textwidth}}
   }
\hbox{\footnotesize \hspace{0.5in} (a) intended receiver
\hspace{0.8in} (b) eavesdropper} \caption{Code rate $R_1$ versus
$R_2$ for the intended receiver and eavesdropper.} \label{fig:macs}
\vspace{-0.2cm}
\end{figure}
However, the receiver cares about only $\Cc_1$ and does not need to
decode $\Cc_2$. Hence, as shown in Fig.~\ref{fig:macs}, the
``achievable'' rate region in the $R_1$-$R_2$ plane at the receiver
is the union of $\Rc^{[\rm MAC]}_1$ and $\Rc^{[\rm S]}_1$. Here
$\Rc^{[\rm MAC]}_1$ is the capacity region of the MAC $(\Xc_1,\Xc_2)
\rightarrow \Yc_1$, in which the intended receiver can decode both
$\Cc_1$ and $\Cc_2$, while $\Rc^{[\rm S]}_1$ is the region in which
the receiver treats codewords from $X_2$ as noise and decodes
$\Cc_1$ only.
Similar analysis applies for the eavesdropper as shown in
Fig.~\ref{fig:macs}.b.
We note that a
proper choice of the redundancy rate $R_2$ can put the eavesdropper
in its unfavorable condition, which can increase secrecy. In the
following, we consider three typical cases: very strong
interference, strong interference, and weak interference. The
analysis for general cases can be found in \cite{Tang:Preprint:07}.

\subsection{Very Strong Interference}
\begin{figure}[htb]
 \centerline{\includegraphics[width=0.4\linewidth,draft=false]{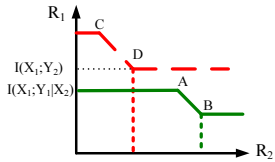}}
  \caption{Very strong interference channel}
  \label{fig:IC-VS}
  \vspace{-0.4cm}
\end{figure}

Fig.~\ref{fig:IC-VS} illustrates the interference channel with very
strong interference. In this case, since
\begin{align}
I(X_1;Y_2)\ge I(X_1;Y_1|X_2),
\end{align}
we cannot obtain any positive secrecy rate.

\subsection{Strong Interference}
We consider strong interference, i.e.,
\begin{align}
&  &        I(X_1;Y_1|X_2)&\le I(X_1;Y_2|X_2) &\notag\\
&\text{and}& I(X_2;Y_2|X_1)& \le I(X_2;Y_1|X_1) &\label{eq:IC-S}
\end{align}
for all product distributions on the input $X_1$ and $X_2$. This
condition implies that, without the interferer, channel
$\Xc_1\rightarrow \Yc_2$ is more capable than channel
$\Xc_1\rightarrow \Yc_1$ and, hence, the achievable secrecy rate may
be $0$.

\begin{figure}[htbp]
  \centerline{\hbox{ \hspace{0.01in}
    \epsfig{file=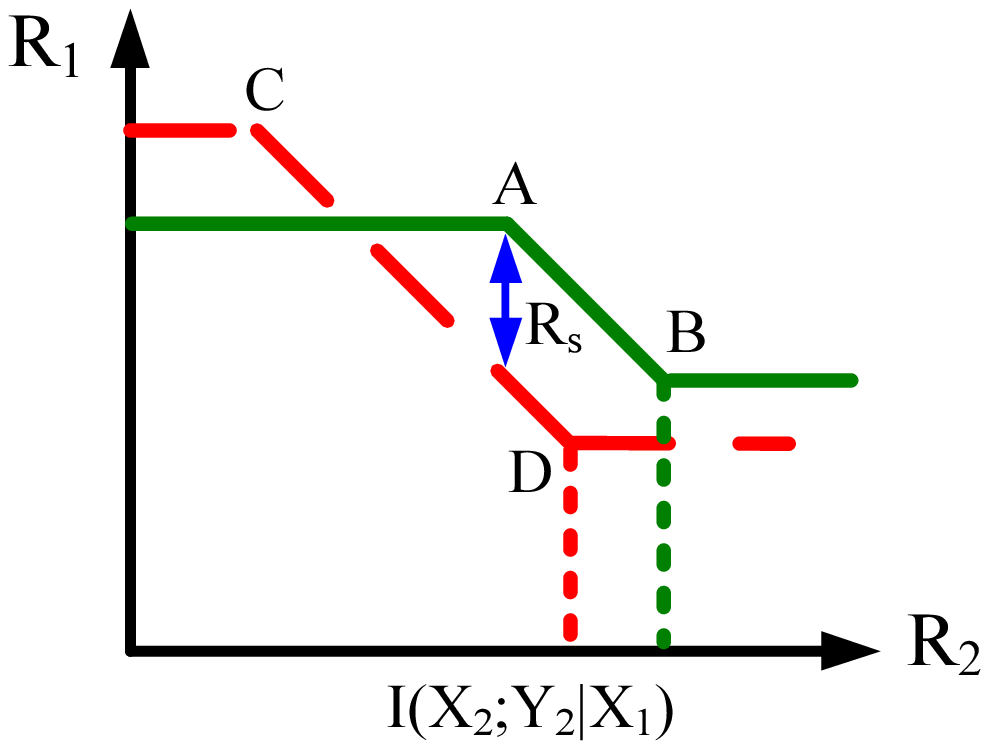, angle=0, width=0.18\textwidth}
    \epsfig{file=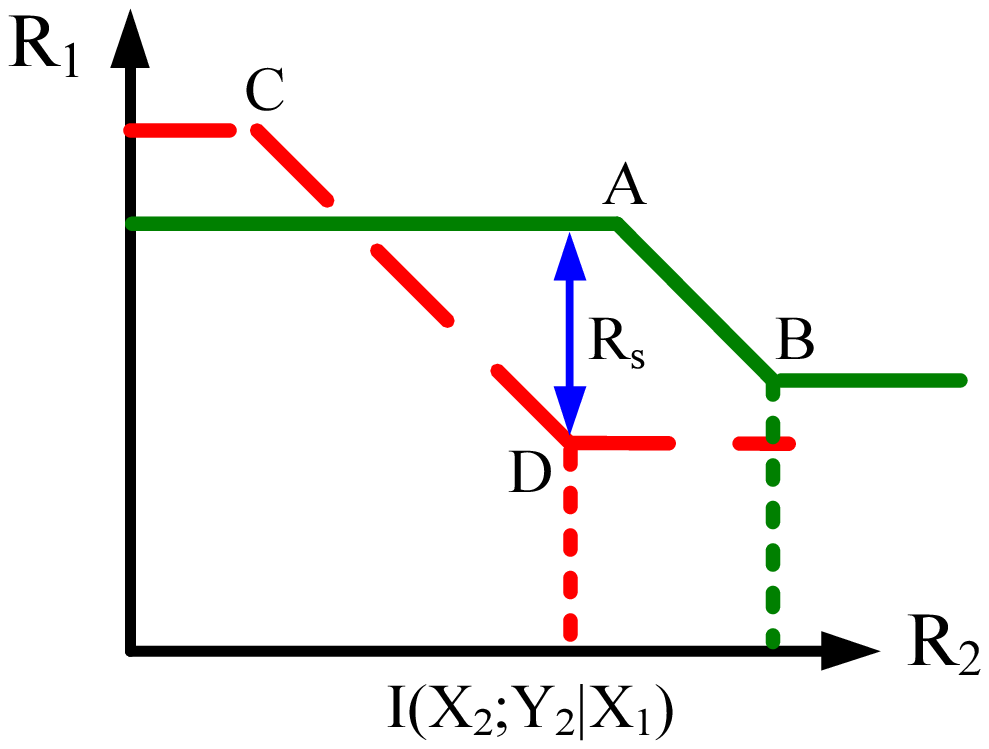, angle=0, width=0.18\textwidth}
    }
  }
    \hbox{\footnotesize \hspace{0.1in} (a) $I(X_2;Y_1)\le I(X_2;Y_2|X_1)$ \hspace{0.2in} (b)  $I(X_2;Y_1)> I(X_2;Y_2|X_1)$}
  \caption{Strong interference channel and $I(X_1,X_2;Y_1)>I(X_1,X_2;Y_2)$}
  \label{fig:IC-S}
  \vspace{-0.2cm}
\end{figure}
However, as shown in Fig.~\ref{fig:IC-S}, we may achieve a positive
secrecy rate with the help of the interferer. Here we choose the
rate pair $(R_1,R_2)\in \Rc_1^{[\rm MAC]}$ so that the intended
receiver can first decode $\Cc_2$ and then $\Cc_1$. Moreover, the
dummy rate pair satisfies
$$(R_{1,d},R_2)\notin \left\{\Rc_2^{[\rm MAC]}\cup \Rc_2^{[\rm S]}\right\},$$
i.e., we provide enough randomness to confuse the eavesdropper. Hence, for
strong interference, the achievable secrecy rate can be simplified as
\begin{align*}
R_s = \max_{\pi}\left\{ \min \left[
               \begin{array}{l}
                I(X_1,X_2;Y_1)-I(X_1,X_2;Y_2),\\
                I(X_1;Y_1|X_2)-I(X_1;Y_2)
               \end{array}
             \right]\right\}^{+}.
\end{align*}

\subsection{Weak Interference}
\begin{figure}[htbp]
  \centerline{\hbox{ \hspace{0.01in}
    \epsfig{file=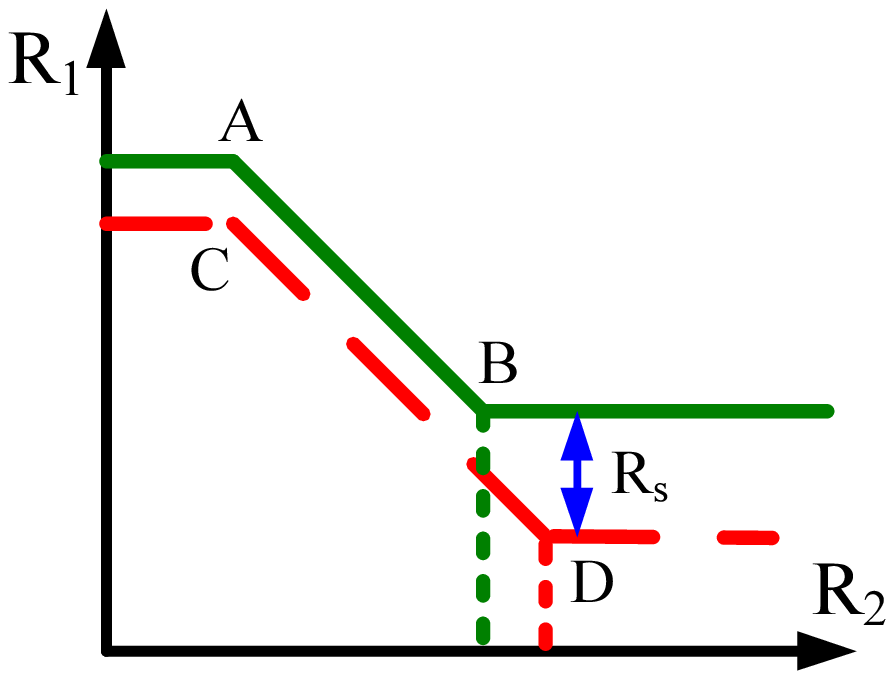, angle=0, width=0.18\textwidth}
    \epsfig{file=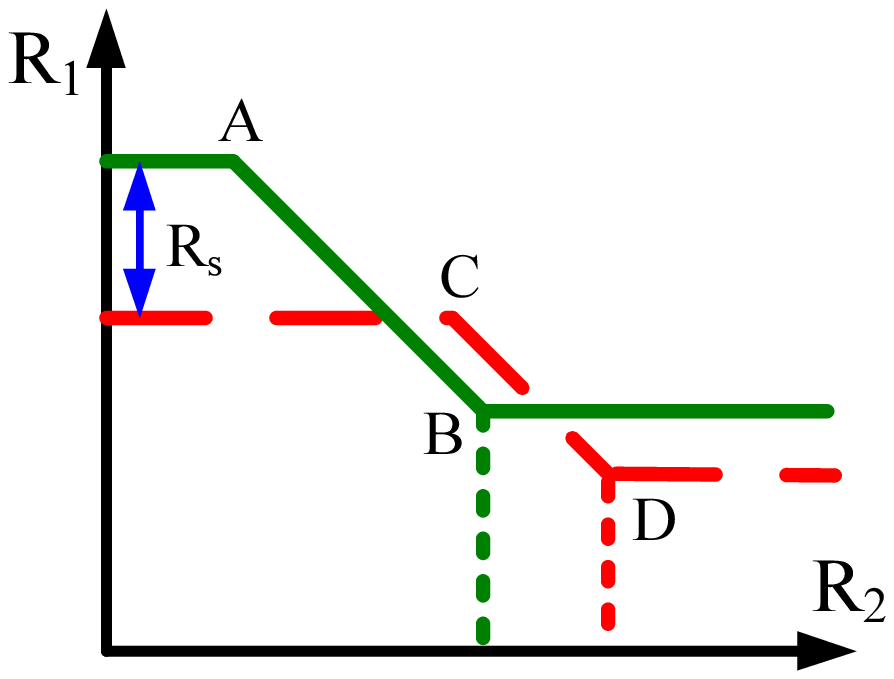, angle=0, width=0.18\textwidth}
    }
  }
    \hbox{\footnotesize \hspace{0.7in} (a) $\Delta_1 \le \Delta_2$ \hspace{0.8in} (b)  $\Delta_1 > \Delta_2$}
  \caption{Weak interference channel}
  \label{fig:IC-W}
  \vspace{-0.2cm}
\end{figure}

Weak interference implies that
\begin{align}
&  &         I(X_1;Y_1|X_2)&\ge I(X_1;Y_2|X_2)&\notag\\
&\text{and}& I(X_2;Y_2|X_1)&\ge I(X_2;Y_1|X_1) & \label{eq:IC-W}
\end{align}
for all product distributions on the input $X_1$ and $X_2$. Let
\begin{align}
&  &   \Delta_1&=I(X_1;Y_1|X_2)-I(X_1;Y_2|X_2)&\\
&\text{and}& \Delta_2&=I(X_1;Y_1)-I(X_1;Y_2). &
\end{align}
As shown in Fig.~\ref{fig:IC-W}.a, the achievable secrecy can be
increased by the help from the interferer when
$\Delta_1\le\Delta_2$.
In this case, the interferer generates an ``artificial noise'' with
the dummy rate $R_2>I(X_2;Y_2|X_1)$ so that neither the receiver nor
the eavesdropper can decode $\Cc_2$. On the other hand, when
$\Delta_1>\Delta_2$, the interferer ``facilitates'' the transmitter
by properly choosing the signal $X_2$ to maximize $\Delta_1$. In the
case of weak interference, the achievable secrecy rate can be
summarized as
\begin{align*}
R_s = \max_{\pi}\left\{ \max \left(\Delta_1,\Delta_2\right)\right\}.
\end{align*}

\section{Symmetric Gaussian Channels}\label{sec:Gaussian}

\begin{figure} [hbt]
  \centering
  \includegraphics[width=2.6in]{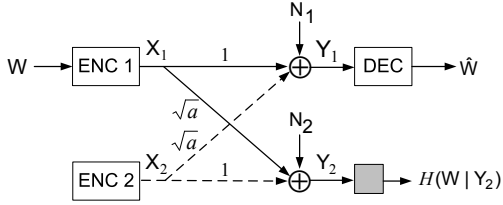}\\
  \caption{A symmetric Gaussian wiretap channel with a helping interferer.}\label{channel}
  \vspace{-0.2cm}
\end{figure}

In this section, we consider the Gaussian wiretap channel with a
helping interferer (GWT-HI). In order to introduce the results in
the simplest possible setting, in this paper we focus on a symmetric
Gaussian channel as illustrated in Fig.~\ref{channel}, where the
source-eavesdropper and interferer-receiver channels have the same
channel condition. The results for the GWT-HI with general parameter
settings can be found in \cite{Tang:Preprint:07}.

The channel outputs at the intended receiver and the eavesdropper
can be written as
\begin{eqnarray}\label{signal}
  Y_{1,k} &=& X_{1,k} +\sqrt{a}X_{2,k} + N_{1,k},  \nonumber\\
  Y_{2,k} &=& \sqrt{a}X_{1,k} + X_{2,k} + N_{2,k},
\end{eqnarray}
for $k=1, \dots, n$, where ${N_{1,k}}$ and ${N_{2,k}}$ are sequences
of independent and identically distributed zero-mean Gaussian noise
variables with unit variances. The channel inputs $X_{1,k}$ and
$X_{2,k}$ satisfy average block power constraints of the form
\begin{equation}\label{power}
  \frac{1}{n}\sum_{k=1}^{n}E[X_{1,k}^2] \leq \bar{P_1}, \quad \frac{1}{n}\sum_{k=1}^{n}E[X_{2,k}^2] \leq \bar{P_2},
\end{equation}

\subsection{Achievable Secrecy Rate}

We give an achievable secrecy rate by assuming that both encoders
use Gaussian codebooks. In this subsection, we assume that the
codewords in $\Cc_1$ and $\Cc_2$ have average block powers $P_1$ and
$P_2$, respectively. The optimal $P_1$ and $P_2$ satisfying the
requirements of $P_1 \leq \bar{P_1}$ and $P_2 \leq \bar{P_2}$ are
found in Subsection \ref{sec:power}.

\begin{theorem}
For the symmetric Gaussian wiretap channel with a helping interferer
given by (\ref{signal}),

i) if $a \geq 1+P_2$, the achievable secrecy rate is $R_s=0$;

ii) if $1 \leq a < 1+P_2$, the achievable secrecy rate is
\begin{eqnarray}\label{rate1}
 \lefteqn{R_s(P_1, P_2) = } \nonumber\\
 &\left\{ \begin{array}{ll}
  \g(P_1) - \g(\frac{aP_1}{1+P_2})  &\mbox{if $P_1<P_2$, $a > 1 +P_1 $,}\\
  \g(P_1+aP_2) - \g(aP_1+P_2) &\mbox{if $P_1 < P_2$, $a \leq 1 + P_1$,}\\
  0  &\mbox{otherwise;}
  \end{array} \right.\nonumber
\end{eqnarray}

iii) if $a<1$, the achievable secrecy rate is
\begin{eqnarray}\label{rate2}
R_s(P_1, P_2) =  \left\{ \begin{array}{ll}
  \g(\frac{P_1}{1+aP_2}) - \g(\frac{aP_1}{1+P_2})  &\mbox{if $P_1 > P_2$,}\\
  \g(P_1) - g(aP_1) &\mbox{otherwise,}
  \end{array} \right.\nonumber
\end{eqnarray}
where $\g(x)=(1/2)\log_2(1+x)$.
\end{theorem}

\begin{proof}
We use the achievability scheme in Theorem$~1$ with Gaussian input
distributions.
\end{proof}

\begin{remark}
For comparison, we recall that the secrecy capacity of the Gaussian
wiretap channel \cite{Leung-Yan-Cheong:IT:78} (the case without an interferer in the GWT-HI model)
is
\begin{eqnarray}\label{gwiretap}
  R_s^{\mathrm{WT}}(P_1)=\left\{ \begin{array}{ll}
  \g(P_1)-\g(aP_1) &\mbox{if $a<1$,}\\
  0  &\mbox{if $a \geq 1$.}
  \end{array} \right.
\end{eqnarray}
That is, a positive secrecy rate can be achieved for the wiretap
channel only when $a<1$. According to Theorem$~2$, a positive
secrecy rate can be achieved for the symmetric GWT-HI when
$a<1+P_2$. If the interferer has sufficiently large power, a
positive secrecy rate can be achieved for any $a>0$.

\end{remark}

\begin{remark}
$a \geq 1+P_2$, $1 \leq a < 1+P_2$, and $a<1$ fall into the cases of
very strong interference, strong interference and weak interference,
respectively.
\end{remark}

\subsection{Power Control}\label{sec:power}

Power control is essential to interference management for
accommodating multi-user communications. As for the GWT-HI, power
control also plays a critical role. In this subsection, we consider
the optimal power control strategy for increasing the secrecy rate
given in Theorem$~2$.

\begin{theorem}
When $a \geq 1$, the power control scheme for maximizing the secrecy
rate is given by
\begin{eqnarray}\label{power1}
(P_1, P_2) = \left\{ \begin{array}{ll}
  \left(\min\{\bar{P_1},P_1^{\ast}\}, \bar{P_2}\right)  &\mbox{if $\bar{P_2} > a-1 $,}\\
  (0,0)  &\mbox{otherwise,}
  \end{array} \right.
\end{eqnarray}
where $P_1^{\ast}=a-1$.

When $a < 1$, the power control scheme for maximizing the secrecy
rate is given by
\begin{equation}\label{power2}
(P_1, P_2) =  \left(\bar{P_1}, \min\{\bar{P_2}, P_2^{\ast}\}\right),
\end{equation}
where
\begin{equation}\label{past}
    P_2^{\ast}=\frac{\sqrt{1+(1+a)\bar{P_1}}-1}{1+a}.
\end{equation}
\end{theorem}

\begin{proof}
The proof can be found in \cite{Tang:Preprint:07}.
\end{proof}

\begin{remark}
When $a<1$, the interferer controls its power so that it does not bring too much interference to the primary transmission. When $a \geq 1$, the benefits of power control at the transmitter are two-fold: First, less information is leaked to the eavesdropper; and furthermore, the intended receiver can successfully decode (and
cancel) the interference.
\end{remark}

\subsection{Power-Unconstrained Secrecy Rate}\label{sec:pcon}

A fundamental parameter of wiretap-channel-based wireless secrecy systems is the achievable secrecy rate when the transmitter has unconstrained power. This secrecy rate is related only to the channel conditions, and is the maximal achievable secrecy rate no matter how large the transmit power is. For example, the power-unconstrained secrecy rate for a Gaussian wiretap channel (when there is
no interferer in the GWT-HI model) is given by
\begin{equation}\label{limit1}
    \lim_{\bar{P_1}\rightarrow
\infty}R_s^{\mathrm{WT}}(\bar{P_1})=\lim_{\bar{P_1} \rightarrow
\infty}\left[\g(\bar{P_1})-\g(a\bar{P_1})\right]^{+}=\frac{1}{2}\left[\log_{2}\frac{1}{a}\right]^{+}.
\end{equation}

After some limiting analysis, we have the following result for the
symmetric GWT-HI model.

\begin{theorem}
The achievable power-unconstrained secrecy rate for the symmetric
GWT-HI is
\begin{eqnarray}\label{limit2}
\lim_{\bar{P_1},\bar{P_2} \rightarrow \infty}R_s = \left\{
\begin{array}{ll}
  \frac{1}{2}\log_{2}a  &\mbox{if $a \geq 1$,}\\
  \log_{2}\frac{1}{a}  &\mbox{if $a < 1$.}
  \end{array} \right.
\end{eqnarray}
\end{theorem}

\begin{proof}
The proof can be found in \cite{Tang:Preprint:07}.
\end{proof}

When the interference is strong ($a>1$), the power unconstrained secrecy rate is $(1/2)\log_{2}a$. Note that $(1/2)\log_{2}a$ is the power-unconstrained secrecy rate if confidential messages are sent from the interferer to the intended receiver in the presence of the eavesdropper. This is particularly interesting because we do not even assume that there is a source-interferer channel (which enables the interferer to relay the transmission). When the interference is weak ($a<1$), the interferer assists the secret transmission by
doubling the achievable secrecy rate.

\section{Numerical Examples}\label{sec:numerical}

\begin{figure}
  \centering
  \includegraphics[width=2.1in]{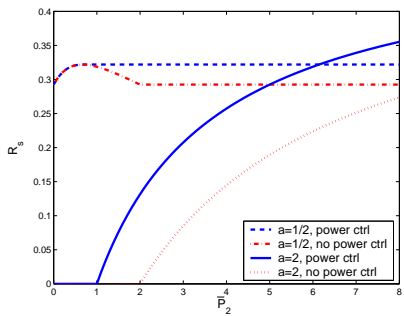}\\
  \caption{Secrecy rate $R_s$ versus $\bar{P_2}$, where $\bar{P_1}=2.$}\label{intpower}
  \vspace{-0.3cm}
\end{figure}

In Fig. \ref{intpower}, we present a numerical example to show the benefits of the power control scheme to the secrecy rate $R_s$. In this example, we assume that the source power constraint is $\bar{P_1}=2$, and the interferer power constraint $\bar{P_2}$ varies from $0$ to $8$. We can see that the power control scheme can increase the secrecy rate significantly. When $a=2$, the power control scheme uses the maximum interferer power and holds the source power to be $P_1^{\ast}=1$, so that the intended receiver can decode the interference first. When $a=1/2$, the power control scheme uses the maximum source power and holds the interferer power below $P_2^{\ast}=2/3$, so that the interferer does not introduce too much interference to the intended receiver (which treats the
interference as noise in this case).

\begin{figure}
  \centering
  \includegraphics[width=2.1in]{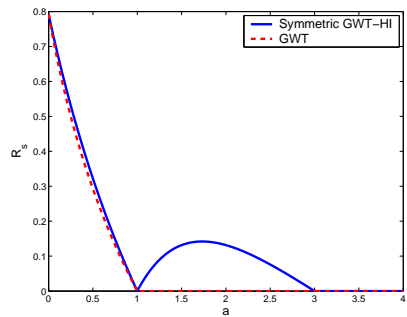}\\
  \caption{Secrecy rate $R_s$ versus $a$, where $\bar{P_1}=\bar{P_2}=2$.}\label{chgain}
  \vspace{-0.3cm}
\end{figure}

In Fig. \ref{chgain}, we present another example to show the achievable secrecy rate $R_s$ for different values of $a$. In this example, we assume that $\bar{P_1}=\bar{P_2}=2$, and $a$ varies from $0$ to $4$. Comparing the secrecy rates achievable for the GWT-HI and GWT, we find that an independent interferer increases $R_s$. For the GWT, $R_s$ decreases with $a$ and remain $0$ when $a \geq 1$. For the GWT-HI, $R_s$ first decreases with $a$ when $a<1$; when $1<a \leq 1.73$, $R_s$ increases with $a$ because the intended receiver now can decode and cancel the interference, while the eavesdropper can only treats the interference as noise; when $a>1.73$, $R_s$ decreases again with $a$ because the interference does not hurt the eavesdropper much when $a$ is large. In particular, when $a \geq 3 (=1+\bar{P_2})$, the eavesdropper can fully decode the primary transmission by treating the interference as noise. Therefore,
$R_s=0$ when $a \geq 3$.


\section{Conclusions}\label{sec:conclusions}
In this paper, we have considered the use of the superposition property of the wireless medium to alleviate the eavesdropping issues caused by the broadcast nature of the medium. We have studied a wiretap channel with a helping interferer, in which the interferer assists the secret communication by injecting independent interference. We have given an achievable secrecy rate with its geometrical interpretation. The results show that interference can be exploited to benefit secret wireless communication.



\bibliographystyle{IEEEtran}
\bibliography{MacFC}

\begin{thebibliography}{10}
\providecommand{\url}[1]{#1}
\csname url@samestyle\endcsname
\providecommand{\newblock}{\relax}
\providecommand{\bibinfo}[2]{#2}
\providecommand{\BIBentrySTDinterwordspacing}{\spaceskip=0pt\relax}
\providecommand{\BIBentryALTinterwordstretchfactor}{4}
\providecommand{\BIBentryALTinterwordspacing}{\spaceskip=\fontdimen2\font plus
\BIBentryALTinterwordstretchfactor\fontdimen3\font minus
  \fontdimen4\font\relax}
\providecommand{\BIBforeignlanguage}[2]{{%
\expandafter\ifx\csname l@#1\endcsname\relax
\typeout{** WARNING: IEEEtran.bst: No hyphenation pattern has been}%
\typeout{** loaded for the language `#1'. Using the pattern for}%
\typeout{** the default language instead.}%
\else
\language=\csname l@#1\endcsname
\fi
#2}}
\providecommand{\BIBdecl}{\relax}
\BIBdecl

\bibitem{Wyner:BSTJ:75}
A.~D. Wyner, ``The wire-tap channel,'' \emph{Bell Syst. Tech. J.}, vol.~54,
  no.~8, pp. 1355--1387, Oct. 1975.

\bibitem{Csiszar:IT:78}
I.~Csisz{\'{a}}r and J.~K{\"{o}}rner, ``Broadcast channels with confidential
  messages,'' \emph{IEEE Trans. Inf. Theory}, vol.~24, no.~3, pp. 339--348, May
  1978.

\bibitem{Leung-Yan-Cheong:IT:78}
S.~K. Leung-Yan-Cheong and M.~Hellman, ``The {G}aussian wire-tap channel,''
  \emph{IEEE Trans. Inf. Theory}, vol.~24, no.~4, pp. 451--456, July 1978.

\bibitem{Liang:IT:06}
Y.~Liang and H.~V. Poor, ``Multiple access channels with confidential
  messages,'' \emph{IEEE Trans. Inf. Theory}, vol.~54, no.~3, pp. 976--1002,
  Mar. 2008.

\bibitem{Liu:ISIT:06}
R.~Liu, I.~Maric, R.~Yates, and P.~Spasojevi\'{c}, ``The discrete memoryless
  multiple access channel with confidential messages,'' in \emph{Proc.\ IEEE
  Int. Symp. Information Theory}, Seattle, WA, USA, July 2006.

\bibitem{Tekin:IT:06}
E.~Tekin and A.~Yener, ``The {G}aussian multiple access wire-tap channel,''
  \emph{IEEE Trans. Inf. Theory}, May 2006, submitted.

\bibitem{Tang:ITW:07}
X.~Tang, R.~Liu, P.~Spasojevi\'{c}, and H.~V. Poor, ``Multiple access channels
  with generalized feedback and confidential messages,'' in \emph{Proc.\ IEEE
  Inf. Theory Workshop}, Lake Tahoe, CA, USA, Sept. 2007.

\bibitem{Tekin:IT:07}
E.~Tekin and A.~Yener, ``The general {G}aussian multiple-access and two-way
  wire-tap channels: Achievable rates and cooperative jamming,'' \emph{IEEE
  Trans. Inf. Theory}, vol.~54, no.~6, Jun. 2008, to appear.

\bibitem{Liu:IT:07}
R.~Liu, I.~Maric, P.~Spasojevi\'{c}, and R.~Yates, ``Discrete memoryless
  interference and broadcast channels with confidential messages: Secrecy
  capacity regions,'' \emph{IEEE Trans. Inf. Theory}, vol.~54, no.~6, Jun.
  2008, to appear.

\bibitem{Liang:Allerton:07}
Y.~Liang, A.~Somekh-Baruch, H.~V. Poor, S.~Shamai, and S.~Verd{\'{u}},
  ``Cognitive interference channels with confidential messages,'' in
  \emph{Proc.\ 45th Annual Allerton Conference on Commun.\, Contr.\,
  Computing}, Monticello, IL, USA, Sept. 2007.

\bibitem{Lai:IT:06}
L.~Lai and H.~{El~Gamal}, ``The relay-eavesdropper channel: Cooperation for
  secrecy,'' \emph{IEEE Trans. Inf. Theory}, Dec. 2006, submitted.

\bibitem{Yusel:CISS:07}
M.~Yuksel and E.~Erkip, ``The relay channel with a wire-tapper,'' in
  \emph{Proc.\ 41st Annual Conference on Information Sciences and Systems},
  Baltimore, MD, Mar. 2007.

\bibitem{Tang:Preprint:07}
X.~Tang, R.~Liu, P.~Spasojevi\'{c}, and H.~V. Poor, ``Interference-assisted
  secret communication,'' in preparation.

\end{thebibliography}

\end{document}